\documentclass{article}
\usepackage{spconf,amsmath,graphicx}
\usepackage{amssymb}
\usepackage{amsmath}
\usepackage{amsthm}
\usepackage{graphicx}
\usepackage{subfigure, epsfig}
\usepackage{pstricks}
\usepackage{pstricks-add}
\usepackage{pst-plot}
\usepackage{bm}
\usepackage{array}
\usepackage{tikz}
\usetikzlibrary{plotmarks}
\usetikzlibrary{shapes,arrows}
\usepackage{bbold}
\usepackage[ruled]{algorithm2e}
\newtheorem{prop}{Proposition}

\title{Crawford-Sobel meet Lloyd-Max on the grid}
\name{B. Larrousse, O.  Beaude, and S. Lasaulce}
   \address{L2S (CNRS-Sup\'{e}lec-Univ. Paris Sud 11),
   3 Rue Joliot-Curie, 91192 Gif-sur-Yvette, France \\
   \{ larrousse, beaude, lasaulce \}  @lss.supelec.fr}
\begin{document}
%
\maketitle

\begin{abstract}
The main contribution of this work is twofold. First, we apply, for the first time, a framework borrowed from economics to a problem in the smart grid namely, the design of signaling schemes between a consumer and an electricity aggregator when these have non-aligned objectives. The consumer's objective is to meet its need in terms of power and send a request (a message) to the aggregator which does not correspond, in general, to its actual need. The aggregator, which receives this request, not only wants to satisfy it but also wants to manage the cost induced by the residential electricity distribution network. Second, we establish connections between the exploited framework and the quantization problem. Although the model assumed for the payoff functions for the consumer and aggregator is quite simple, it allows one to extract insights of practical interest from the analysis conducted. This allows us to establish a direct connection with quantization, and more importantly, to open a much more general challenge for source and channel coding.
\end{abstract}

\section{Introduction}
\label{sec:introduction}

In today's electricity networks, the energy production is mainly driven by the consumer's demand. However, in the smart grid it will be more common that the consumer will have to adapt its consumption to production e.g., when an erasure mechanism is implemented or when the energy source is a solar/wind farm. Obviously, the consumer and aggregator (i.e., the entity which takes the decision to which extent to meet the demand) will have diverging objectives in general. As a consequence, it might happen that the consumer reports a demand which is higher than the actual need to be effectively satisfied. As, in practice, the request of the consumer (a factory, a house, an EV's owner, etc.) in terms of needed power and the decision to which extent to deliver it by the aggregator (a utility company, a distribution network's operator, etc) will quite often result from (automated) procedures implemented by machines, an important engineering problem appears: How to design a point-to-point communication system where the transmitter (or coder) and receiver (or decoder) have diverging objectives? Indeed, the classical paradigm in communication systems \cite{Shannon1948}, assumes that the coder and decoder have a common objective (e.g., to minimize the distortion or block error rate). When the coder and decoder have non-aligned objectives, the problem of (source/channel) coding needs to be revisited. In the present paper, we will only make a small step into the direction of answering the aforementioned fundamental question. Nonetheless, the work reported here has the merit to bridge an obvious gap between the economics literature and the one of signal processing and communications.

Specifically, we consider an aggregator whose objective is to satisfy the consumer but also to minimize the operating cost induced by the distribution network. More precisely, the cost of the distribution network is chosen to be the residential transformer ageing\footnote{Note that, in France, for instance, there are about 700 000 residential transformers, which shows the importance of managing transformers ageing; see e.g., \cite{Gong2011}\cite{McBee2009}\cite{Tran2012} for more motivations.}. On the other hand, the consumer's ultimate objective is to obtain an amount of power (or energy) as close as possible to its actual need. Based on a signal/message received from the consumer about its need in terms of power, the aggregator eventually decides the amount of power effectively allocated to the consumer. One of the purposes of this paper is to construct a signaling scheme from the consumer to the aggregator which would allow them to reach a consensus or equilibrium about how to communicate in practice (based on a suited communication standard). It turns out that, by considering a simple but realistic model for the aggregator and consumer costs, the problem to be solved is a game whose formulation is related to the problem of strategic information transmission in  economics \cite{crawford1982} and the one of quantization. Indeed, the problem of strategic information transmission has been introduced in \cite{crawford1982} and developed in economics\footnote{A typical example in economics is the interaction between a recruiting officer and a job seeker (the latter has to reveal more or less information about his state, i.e. his capabilities or expectations in terms of salary, while the former has to decide about the salary level).} (see e.g., \cite{Sobel2007} for a recent survey) but not penetrated engineering yet up to a few exceptions \cite{Kavitha2012}\cite{Meriaux2012}, which do not consider neither the smart grid application nor the connections with coding/quantization.\\
 The paper is organized as follows. The problem is formulated in Sec. \ref{sec:problem-statement}. The signaling scheme is determined in Sec. \ref{sec:signaling-scheme} as well as its main properties of practical interest. Sec. \ref{sec:numerical-analysis} provides numerical results to illustrate the derivations of Sec. \ref{sec:signaling-scheme}. Sec. \ref{sec:conclusion} concludes the paper. \\


\section{Problem formulation}
\label{sec:problem-statement}

Fig.~1 provides several key aspects of the considered problem. We consider a consumer whose objective is to obtain an allocated power which is as close as possible to a desired level denoted by $s \geq 0$. For this purpose, the consumer sends a message $m \in \{1,2,...,M\}$ ($M<+\infty$) to the aggregator through a perfect communication channel. Based on the received message, the aggregator effectively provides an amount of power which is denoted by $a \geq 0$. Without loss of generality, it is assumed that $(a,s) \in [0,1]^2$. One way of mathematically formulating the objective of the consumer, is to consider that he aims at maximizing the following payoff function

\begin{equation}
u_{\text{C}}(s, a) = -(s - a)^2 + K
\label{eq:utility-EV}
\end{equation}
where $K\in \mathbb{R}$ is a constant. With such a model, the consumer both aims at meeting its need in terms of power but also at not exceeding the desired power level, which might for instance induce some unnecessary monetary expenses. This model can also be very well justified when $s$ is interpreted as a desired quantity of energy e.g., for recharging a battery (see, e.g., \cite{Marano2009}). Note that here, for the sake of simplicity, we implicitly assume that the energy need corresponds to a need in terms of load or power, which is very realistic when the consumer obtains a constant power transfer rate; relaxing this assumption can be considered as a possible extension of this work. On the other hand, the aggregator's payoff function is assumed to be the weighted sum of the consumer's payoff and a payoff function related to an operating cost induced by the grid:
\begin{equation}
\begin{array}{ccc}
u_{\text{A}}(s, a) & = &  u_{\text{C}}(s, a) + u_{\text{grid}}(s,a) \\
    & = & -(s - a)^2 + K - b e^{a}
\end{array}
\label{eq:utility-aggr}
\end{equation}
where $b \geq 0$ represents a weight which translates the importance of the component associated with the grid. More precisely, the grid component represents a good model of the ageing acceleration factor of a (residential) transformer (see e.g., \cite{IEC} which justifies why the ageing is accelerated exponentially when operating above its nominal load). In the context of strategic information transmission in economics \cite{crawford1982}, the parameter $b$ is interpreted as a bias which quantifies the divergence of interests between the decision-makers which are the consumer and aggregator here. 

\begin{figure}[!ht]\label{fig:EVsignaling}
\begin{center}
\psset{xunit=0.7cm,yunit=0.7cm}
\begin{pspicture}(0,0)(10,6)
\epsfxsize=2cm
\newrgbcolor{darkgreen}{0 0.4 0}
\newrgbcolor{darkblue}{0 0 0.6}
\newrgbcolor{darkred}{0.6 0 0}
\rput(0.4,3){\rnode{A}{\epsfxsize=13mm  \epsfysize=13mm \epsfbox{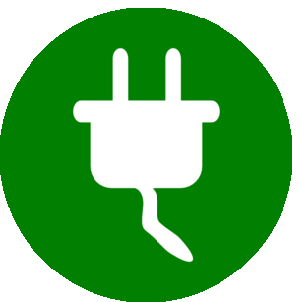}}}
\rput[u](0.6,4.2){\darkgreen Consumer}
\rput[u](0.5,1.7){\darkgreen whose need is $s$}
\rput[u](0.5,1.1){\darkgreen and payoff}
\rput[u](0.5,0.5){\darkgreen $u_{\text{C}}(s,a)$}
\epsfxsize=1.5cm
\rput(9.5,3){\rnode{B}{\epsfbox{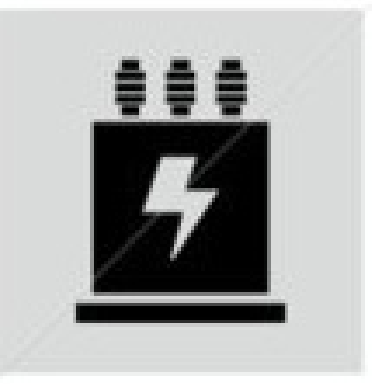}}}
\rput[u](9.5,4.5){Aggregator}
\rput[u](9.5,1.3){with payoff}
\rput[u](9.5,0.7){$u_{\text{A}}(s,a)$}
\psset{nodesep=3pt}
\ncarc[linestyle=dashed,linecolor=darkgreen]{->}{A}{B}
\rput[u](5,4.5){\darkgreen 1. Signaling}
\rput[u](5,3.8){\darkgreen message $m$}
\ncarc[linecolor=gray]{->}{B}{A}
\rput[u](5,2){2. Aggregator's action $a$}
\end{pspicture}
{\small \caption{The consumer (coder) has an actual need in terms of power $s$ which is unknown to the aggregator. The consumer sends a message $m$ to the aggregator (decoder). The aggregator then chooses an action $a$, which is the amount of power effectively allocated to the consumer. The key point is that the aggregator and consumer have \emph{non-aligned} payoff functions.}}
\end{center}
\vspace{-0,5cm}
\end{figure}

One of the contributions of this paper is precisely to inspire from the original framework of \cite{crawford1982} to design a good/consensus/equilibrium signaling scheme between the consumer and aggregator namely, to determine a good signaling scheme in presence of diverging interests between the coder and decoder. First, the consumer should map its knowledge about its actual power need $s$ into the message sent to the aggregator $m$, which amounts to determining a coding function $f$ defined by:
\vspace{-0.2cm}
\begin{equation}
f:\left|
\begin{array}{ccc}
[0, 1] & \rightarrow &  \{1,2,...,M\} \\
 s    & \mapsto & m
\end{array}
\right..
\label{eq:encoding-function}
\end{equation}
Second, the aggregator has to perform the decoding operation by implementing:
\vspace{-0.25cm}
\begin{equation}
g:\left|
\begin{array}{ccc}
\{1,2,...,M\} & \rightarrow &  [0, 1] \\
 m    & \mapsto & a
\end{array}\right..
\label{eq:decoding-function}
\end{equation}
As a first comment note that $f$ and $g$ are deterministic mappings instead of conditional probabilities $q(m|s)$ and $r(a|m)$; this choice does not induce any loss in terms of expected payoff because $u_{\text{A}}$ and $u_{\text{C}}$ are concave. If $b=0$ and the power need $s$ is seen as the realization of a random variable whose distribution $p(s)$ is effectively known to the coder and decoder (this corresponds to a particular scenario in terms of beliefs), the problem of determining $f$ and $g$ can be seen as an instance of a scalar quantization problem which is itself a special case of lossy source coding \cite{Cover2006}. But, in general $b>0$ and, even if the  distribution $p(s)$ is known to both the coder and decoder, the consequence of this simple difference is that the coder, knowing that the decoder has a different objective, will not maximize its expected payoff by revealing its actual need in terms of power. Rather, it will reveal only a degraded version of it and, this, even if $M$ is infinite. As explained in the following section, in general, equilibrium signaling schemes only exploit a fraction of the number of available messages (or bits).

\section{Proposed signaling scheme}
\label{sec:signaling-scheme}

\subsection{Methodology. Connection with quantization}
\label{sec:methodo}

In the presence of decision-makers having different payoff functions and which can only control some variables of the latter, the very meaning of optimality is unclear and the problem needs to be defined before being solved (see e.g., \cite{lasaulce-book-2011}). In this context, one important solution concept is the Nash Equilibrium (NE), which is a vector of strategies from which no decision-maker or player has anything to gain by changing his own strategy unilaterally. Here, we are in the presence of two players namely, the aggregator and consumer. The strategy of the consumer consists in choosing $f$, which corresponds to choosing a partition of the space of possible power needs i.e., $[0,1]$. With each interval is associated a message $m\in \{1,...,M\}$ intended for the aggregator. The strategy of the aggregator consists in choosing $g$ to generate the action $a$, which can be interpreted as choosing a representative of the interval associated with the received message $m$; these intervals are denoted by $I_m = [s_m, s_{m+1}]$. Here, the connection with the quantization problem can be established. Typically, the quantization problem consists in minimizing the distortion $\mathbb{E}\left[(s -\widehat{s})^2 \right]$ ($\widehat{s} = a$ in our setting), with respect to $f$ and $g$. If $f$ and $g$ are optimized separately, the problem can be interpreted as a game where one player chooses $f$ and the other chooses $g$. Since the cost functions are common and the number of message $M$ is fixed, this defines a potential game \cite{Monderer1996}\cite{Lasaulce-Tutorial-09}. In this type of games, it is known that the iterative procedure consisting in optimizing the cost/payoff function w.r.t. $f$ for a fixed $g$, then to optimize it w.r.t. $g$ for the updated $f$, and so on, converges to an NE. This procedure is called the sequential best-response (BR) dynamics in game theory, the BR of a player being the set-valued function which provides the set of strategies which maximize the payoff of this player for a given strategy for the other. The Lloyd-Max algorithm precisely implements this procedure and converges to an NE. Indeed, the intersection points between the players' BRs are precisely the NE of the game. In the next section, we determine the BRs in the considered setting in which players have different payoff functions.

\subsection{Aggregator's best-response} \label{sec:aggreg-BR}

When the aggregator receives a message $m$, it knows that the actual consumer's power need $s$ is in the interval $I_m $ but not its exact value. Therefore, in general, given the knowledge of the message, the aggregator has a certain belief about the power need, which is denoted by $\pi_{\text{A}}(s|m)$. The aggregator best-responds to the message by maximizing the expected payoff that is,
\begin{equation}
a^{\star}(m) = \int_0^1
u_{\text{A}}(a,s) \pi_{\text{A}}(s|m) \mathrm{d}s.
\end{equation}

Here, we assume that this belief is chosen to be a uniform probability distribution over the interval $I_m$ which corresponds to the case where the aggregator has no statistical information at all about the consumer's power need; other scenarios in terms of belief are left as extensions. The following proposition provides the expression of the aggregator BR i.e., the best representative of the interval $I_m$.

\begin{prop} Given a partition scheme $f$ (or $m(s)$), the aggregator's best-response $a^{\star}(m)$ to a message $m$ is:\begin{equation}
\label{eq:CondOpt}
a^{\star}(m) = \left|
\begin{array}{ccl}
\overline{s}_m   - W \left(b \frac{e^{\overline{s}_m}}{2}\right) & \text{if} & \overline{s}_{m} > \frac{b}{2} \\
0 & \text{if} &  \overline{s}_{m} \leq \frac{b}{2} \\
\end{array}
\right.
\end{equation}
where
$\overline{s}_m = \frac{s_m+s_{m+1}}{2}$ and $W$ is the Lambert $W$ function\footnote{ Some basics on the Lamber $W$ function can be found here: http://mathworld.wolfram.com/LambertW-Function.html}.
\end{prop}

The proof of this result is not provided here. It is constructed from arguments exploited in \cite{crawford1982} which uses a useful property (called the single crossing condition) of the payoff functions. This property is as follows:
\begin{equation}\label{eq:single-crossing-condition}
\forall i \in \left\{\text{A},\text{C}\right\},
\left\{
\begin{array}{cc}
\forall s, & \exists a, \frac{\partial u_i}{\partial s}(s,a) = 0 \\
\forall (a,s), & \frac{\partial^2 u_i}{\partial s^2}(a,s) < 0 \\
\forall (a,s), & \frac{\partial^2 u_i}{\partial s \partial a}(a,s) > 0 \\
\end{array}
\right..
\end{equation}
The above result shows that it is possible to express the aggregator's best action (for a given message) in a simple way. The first term of the optimal action corresponds to what is called the centroid in quantization. The presence of the second term is precisely due to the fact that the coder and decoder have diverging interests. In the extreme case where $b\rightarrow0$, the optimal action for the aggregator therefore corresponds to the centroid whereas the optimal action is simply $0$ when $b\rightarrow\infty$.

\subsection{Consumer's best-response}

The consumer's strategy is to choose a partition of the power need space $[0,1]$ into intervals $I_1,I_2,...,I_M$ with $I_m=[s_m,s_{m+1}]$. In contrast with a classical quantization problem, the number of messages (or bits) to be used to form the partition is not fixed and can be optimized by the consumer in order to maximize its expected payoff function for a given action. This feature constitutes an important technical difference which can be incorporated in a non-trivial manner. One key result of \cite{crawford1982} which we re-exploit here is that the optimal partition structure can be built from an optimality condition called the arbitrage condition. Intuitively, if the consumer power need $s$ is exactly $s_m$, then the consumer should be indifferent between sending the messages $m-1$ (associated with the interval $[s_{m-1}, s_m]$) and $m$ (associated with the interval $[s_{m}, s_{m+1}]$). By exploiting this optimality condition and the single crossing condition (\ref{eq:single-crossing-condition}), the following proposition can be proved.

\begin{prop}
\label{CharOptPart} Let $M_b^{\star}$ be the number of optimal partitions. For a given $L\leq M_b^{\star}$, the optimal partition for the consumer can be defined recursively as:
\begin{equation}
\label{RecursCalc}
\left\{
\begin{array}{l}
s^{\star}_0 = 0 \\
\begin{aligned}
s^{\star}_{m+1}=\phi_b(2s^{\star}_{m}-\phi_b^{-1}(s^{\star}_{m-1}
+s^{\star}_{m}))-s^{\star}_{m}&, \\ 1 \leq m \leq L-1 & \end{aligned}\\
s^{\star}_L = 1\\
\end{array}
\right.
\end{equation}
where $\phi_b(x) = 2x + be^x$ and $\phi_b^{-1}$ its inverse function.
\end{prop}

Prop. 1 and 2 completely define the signaling scheme for the problem under investigation. Indeed, when the consumer chooses a partition of the power need space according to Prop. 2 and the aggregator chooses the representatives according to Prop. 1 , we obtain an NE. The number $M_b^\star$ corresponds to the number of NE. This number can be  obtained by choosing a certain $s_1= b + \epsilon$ and determining the partition $s_2(s_1), s_3(s_1), ...$  through $s^{\star}_{m+1}=\phi_b(2s^{\star}_{m}-\phi_b^{-1}(s^{\star}_{m-1}
+s^{\star}_{m}))-s^{\star}_{m}$, and keeping the greatest integer $m$ such that $s^{\star}_{m+1} < 1$. It can be shown \cite{crawford1982} that an NE based on partition exploiting $L \in \{1,2,...,M_b^{\star}\}$ messages is better ex ante for both players than another NE which exploits $L' < L$ messages. If the bias is greater than a threshold $\beta$, $M_b^{\star}=1$, which means that the consumer's message set is a singleton and no information is revealed to the aggregator. This threshold is important in practice because it allows a designer to know under which conditions a given signaling-based power production cannot be implemented. As a last comment on Prop. 1, note that when $b \rightarrow 0$, the recursive equation boils down to $s^{\star}_{m+1} = 2s^{\star}_{m} - s^{\star}_{m-1}$ i.e., $s^{\star}_{m} = \frac{s^{\star}_{m-1} + s^{\star}_{m+1}}{2}$. Therefore, a partition similar to uniform quantization is obtained.

\section{Numerical results}
\label{sec:numerical-analysis}

\begin{figure}
\begin{center}
\includegraphics[scale=0.39]{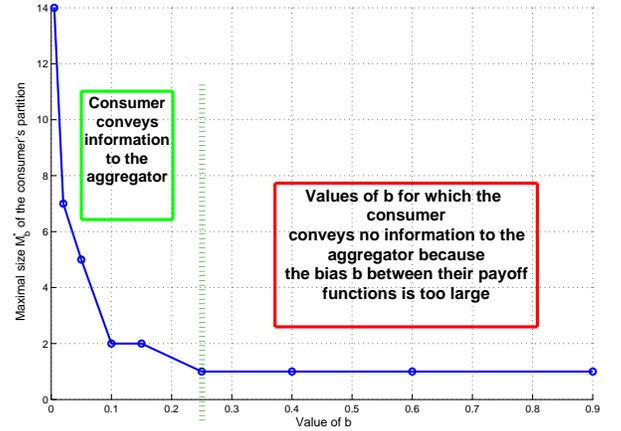}
\caption{Maximal number of messages $M_b^\star$ vs the bias $b$.}
\label{fig:MaxSize}
\end{center}
\end{figure}

\begin{figure}
\includegraphics[scale=0.39]{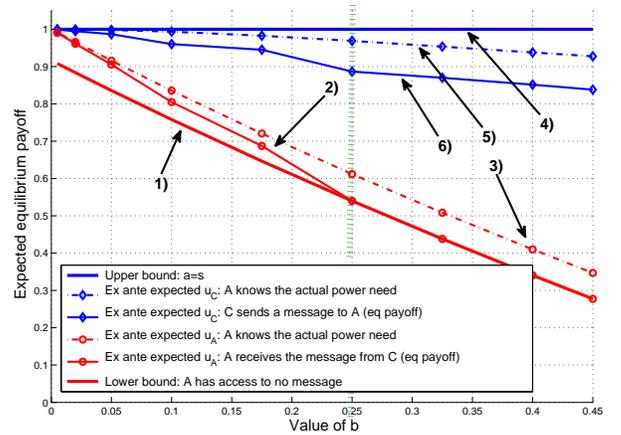}
\caption{Consumer's and aggregator's payoff vs the bias $b$.}
\label{fig:ComparU}
\end{figure}

We assume that the consumer's power need is distributed uniformly over $[0,1]$, $K=1$, and we study the influence of the parameter $b$. In practice, $b$ can be determined by the weight the aggregator puts on the transformer's cost but also physical parameters such as the ambient and hot-spot temperatures, and the nominal load \cite{IEC}. Fig. \ref{fig:MaxSize} quantifies the relation between the bias and the maximum number of messages used at equilibrium (which also corresponds to the number of NE). Here, for $b \geq \beta$, with $\beta \sim 0.25$, the consumer does not reveal anything about its need in terms of power to the aggregator. When $b\rightarrow 0$, the number of messages becomes high and, will be limited, in practice, e.g, by the communication channel capacity. Fig. \ref{fig:ComparU} shows the expected payoff obtained by the aggregator (``A'') and consumer (``C'') as a function of $b$ in different scenarios. The three bottom curves correspond to the aggregator's payoff when: 1) ``A'' has access to no message; 2) ``A'' receives the message from ``C'' (equilibrium payoff); 3) ``A'' is given the perfect knowledge of the power need. The loss induced by the uncertainty on the power need corresponds to the gap between 1) and 3), which may be typically much higher for other payoff functions. The three top curves correspond to the consumer's payoff when: 4) ``C'' $a=s$; 5) ``A'' knows the actual power need; 6) ``C'' sends a message to ``A'' (equilibrium payoff).

\section{Conclusion}
\label{sec:conclusion}

Obviously, the model used in this work can be generalized in many respects e.g., in terms of payoff functions, in terms of beliefs about the source/need distribution, and also by considering sequence of actions instead of a single action. The framework which is exploited in this paper goes beyond the quantization aspects and the new connections established between \cite{crawford1982} and quantization and opens new technical challenges which concern the general problem of source and channel coding when the coders and decoders have different performance criteria.


\bibliographystyle{IEEEbib}
\bibliography{sobelbib2}

\end{document}